\PassOptionsToPackage{svgnames}{xcolor}

\documentclass[twocol]{ametsocV6.1}

\usepackage{graphicx}
\usepackage{caption}
\usepackage[export]{adjustbox}
\usepackage{multirow}
\usepackage{booktabs}
\usepackage{array}
\usepackage{xcolor}

\title{Exploring Randomly Wired Neural Networks for Climate Model Emulation}

\authors{William Yik,\aff{a,b}\correspondingauthor{William Yik, wyik@hmc.edu}
Sam J. Silva, \aff{b,c}
Andrew Geiss, \aff{d}
Duncan Watson-Parris \aff{e,f}}

\affiliation{\aff{a}{Harvey Mudd College, Claremont, CA}\\
\aff{b}{Department of Earth Sciences, University of Southern California, Los Angeles, CA}\\
\aff{c}{Department of Civil and Environmental Engineering, University of Southern California, Los Angeles, CA}\\
\aff{d}{Pacific Northwest National Laboratory, Richland, WA}\\
\aff{e}{Scripps Institution of Oceanography, University of California San Diego, La Jolla, CA}\\
\aff{f}{Hal{\i}c{\i}o\u{g}lu Data Science Institute, University of California San Diego, La Jolla, CA}}

\abstract{Exploring the climate impacts of various anthropogenic emissions scenarios is  key to making informed decisions for climate change mitigation and adaptation. State-of-the-art Earth system models can provide detailed insight into these impacts, but have a large associated computational cost on a per-scenario basis. This large computational burden has driven recent interest in developing cheap machine learning models for the task of climate model emulation. In this manuscript, we explore the efficacy of \textit{randomly wired neural networks} for this task. We describe how they can be constructed and compare them to their standard feedforward counterparts using the ClimateBench dataset. Specifically, we replace the serially connected dense layers in multilayer perceptrons, convolutional neural networks, and convolutional long short-term memory networks with randomly wired dense layers and assess the impact on model performance for models with 1 million and 10 million parameters. We find that models with less complex architectures see the greatest performance improvement with the addition of random wiring (up to 30.4\% for multilayer perceptrons). Furthermore, out of 24 different model architecture, parameter count, and prediction task combinations, only one saw a statistically significant performance deficit in randomly wired networks compared to their standard counterparts, with 14 cases showing statistically significant improvement. We also find no significant difference in prediction speed between networks with standard feedforward dense layers and those with randomly wired layers. These findings indicate that randomly wired neural networks may be suitable direct replacements for traditional dense layers in many standard models.}

\begin{document}

\maketitle

%
%
%
\statement
Modeling various greenhouse gas and aerosol emissions scenarios is important for both understanding climate change and making informed political and economic decisions. However, accomplishing this with large Earth system models is a complex and computationally expensive task. As such, data-driven machine learning models have risen in prevalence as cheap emulators of Earth system models. In this work, we explore a special type of machine learning model called \textit{randomly wired neural networks} and find that they perform competitively for the task of climate model emulation. This indicates that future machine learning models for emulation may significantly benefit from using randomly wired neural networks as opposed to their more standard counterparts.

%








\section{Introduction}\label{sec: intro}
Characterizing the response of the Earth system to future emissions scenarios is key to informing climate change mitigation and adaptation strategies. Modern Earth system models (ESMs) are incredibly useful tools for this task, providing detailed climate projections far into the future for scenario-based analysis. However, as the process complexity and spatial resolution of such ESMs increases, so does their computational cost \citep{collins2012quantifying}. Consequently, it becomes impractical to explore a wide range of possible emissions scenarios with modern ESMs, limiting their applicability to a restricted set of future scenarios \citep{oneill2016scenario}.

To address this large computational cost, the climate modeling research community frequently makes use of emulators. Emulators are a set of statistical tools that aim to approximate the complex physical relationships in a full ESM at a fraction of the computational cost \citep{meinshausen2011emulating,tebaldi2020emulating}. Recent developments have demonstrated the considerable promise of machine learning (ML) techniques in these emulation tasks for both individual model components \citep{seifert2020potential,silva2021physically,mooers2021assessing,chantry2021machine,lee2019estimation} and entire ESM predictions \citep{rasp2020weatherbench,watson2022climatebench,mansfield2020predicting,beusch2020emulating}. Such ML models often enable accurate predictions at greatly reduced computational cost relative to full ESMs. 

The model classes investigated in state-of-the-art ML research in the climate sciences range widely, and include linear models \citep{rasp2020coupled,silva2020development,mansfield2020predicting}, tree based methods \citep{yuval2020stable, yuval2021use, silva2021physically}, and various implementations of neural networks \citep{price2022increasing, bretherton2022correcting, krasnopolsky2005new, rasp2021data}. This is broadly indicative of a very large potential design space for ML model architectures. Here, we investigate ``randomly wired neural networks" in an effort to further explore this architecture design space for climate mode emulation. Randomly wired neural networks are a special class of neural network where components are connected in a random, rather than structured, manner. This is in direct contrast to widely-used feed forward multi-layer perceptron (FFMLP) style architectures, where components are connected in series. At its core, random wiring is a form of neural architecture search (NAS) \citep{xie2019exploring, kasim2021building} that searches a more complete space of connectivity patterns than other common NAS strategies \citep{elsken2019neural} to identify high-performing model architectures. This class of neural network introduces novel  connectivity patterns between layers rather than novel types of layers themselves, and they have demonstrated skill in a variety of domains, including handwriting recognition \citep{gelenbe2016deep}, internet network attack detection \citep{brun2018deep}, and emulation of aerosol optics \citep{geiss2022emulating}. 

In this work, we evaluate the suitability of randomly wired neural networks for climate emulation using the ClimateBench benchmarking dataset. We specifically investigate the use of random wiring to predict future temperature and precipitation statistics. To that end, we compare the performance of randomly wired networks against their serially connected counterparts within three types of neural network models covering a wide range of complexities: multilayer perceptrons (MLPs), convolutional neural networks (CNNs), and convolutional long short-term memory networks (CNN-LSTMs). We find that random network wiring shows competitive results across model architectures and prediction tasks with the greatest improvements (up to 30.4\%) in MLPs. These performance improvements for randomly wired models decreases with increasing model architecture complexity, but out of 24 model architecture, parameter count, and prediction task combinations, only one saw a statistically significant performance deficit for randomly wired models. On the other hand, 14 saw statistically significant performance improvement. Furthermore, we find that randomly wired neural networks take no longer to make predictions than their standard serially connected counterparts. Our work indicates that in many cases, random wiring can serve as direct replacements for traditional feedforward neural networks to improve predictive skill.

The remainder of the paper is structured as follows. In Section \ref{sec: rand nn} we provide the details of how we construct our randomly wired neural networks and highlight their differences with more standard networks. Section \ref{sec: data} describes the ClimateBench dataset and how we use it to train, validate, test, and evaluate our models. Our experimental setup to compare randomly wired neural networks with their conventional counterparts is discussed in Section \ref{sec: exp setup}. Section \ref{sec: results} provides the results of these experiments, with further discussion and analysis in Section \ref{sec: discussion}. Lastly, Section \ref{sec: conclusion} concludes the manuscript with discussion of future work.

\section{Randomly Wired Neural Networks}\label{sec: rand nn}
\begin{figure*}[t]
    \begin{minipage}{.5\textwidth}
        \centering
        \includegraphics[valign=c, width=0.75\linewidth]{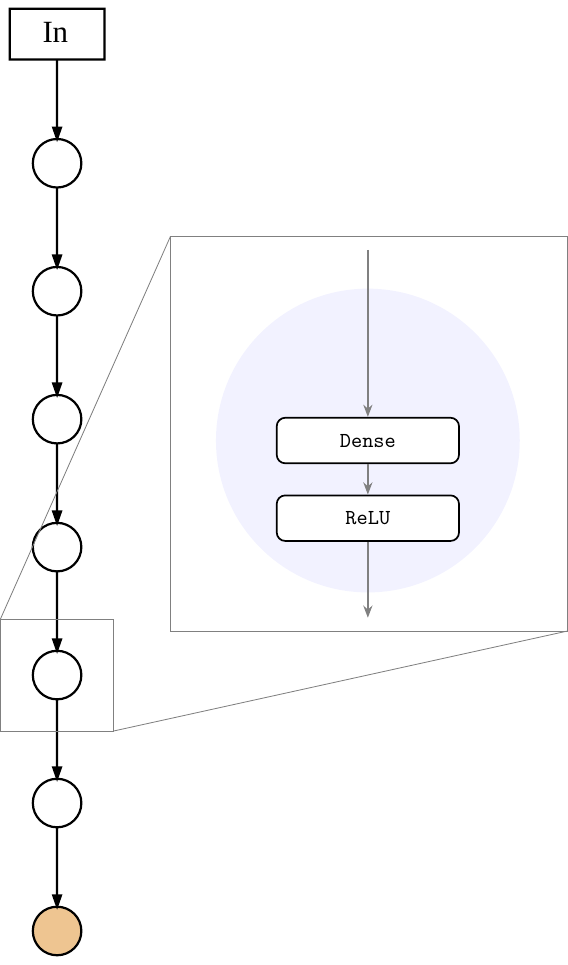}
    \end{minipage}%
    \begin{minipage}{.5\textwidth}
        \centering
        \includegraphics[valign=c, width=0.96\linewidth]{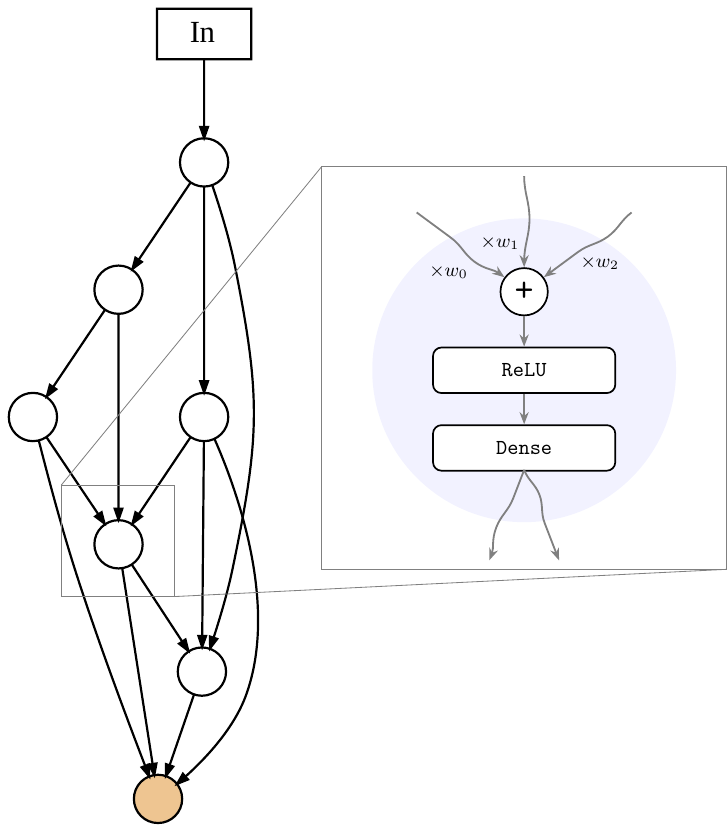}
    \end{minipage}%
    \caption{Graph representations of a six layer MLP (left) and six layer RandDense network (right) shown side-by-side with their respective node operations in the insets. White circles represent hidden dense layers and their activation functions. Brown circles represent the output layer discussed in Section \ref{sec: rand nn}. Lastly, the upper rectangular block represents neural network layers preceding the dense layers. In this case, it is a simple input layer which performs no operation, but other choices such as a convolutional block are possible (see Section \ref{sec: exp setup}). For the RandDense network, aggregation from three previous input nodes is done via weighted sum with weights $w_0$, $w_1$, and $w_2$. The summation is followed by ReLU activation and the dense layer. Lastly, two identical copies of the output are sent to two separate nodes downstream.}
    \label{fig: dense and rand dense}
\end{figure*}

A typical multilayer perceptron (MLP) neural network is structured such that information (a tensor) only flows in a single direction from one dense layer to the next. This is illustrated in Figure \ref{fig: dense and rand dense} which shows a graph representation of an MLP with six hidden dense layers. In this graph representation, each layer is represented as a node in a directed graph where edges indicate the flow of tensors. We use this graph representation and terminology in part for consistency with previous work \citep{xie2019exploring}. In our random networks, as well as those of \citet{xie2019exploring} and \citet{geiss2022emulating}, data still flows through the network in a feedforward manner. However, unlike dense layers in MLPs, dense layers in our random networks may receive inputs from any number of preceding layers and pass their outputs to any number of subsequent layers. That is, there may be ``skip connections" between dense layers. An example of this class of random neural network, which we will henceforth refer to as ``RandDense'' networks, is illustrated in Figure \ref{fig: dense and rand dense}. With this in mind, we will detail our random network generation process and further highlight differences between our networks and typical MLPs.

\begin{figure*}[t]
    \centering
    \begin{minipage}{0.5\textwidth}
        \centering
        \includegraphics[valign=c,width=0.1\linewidth]{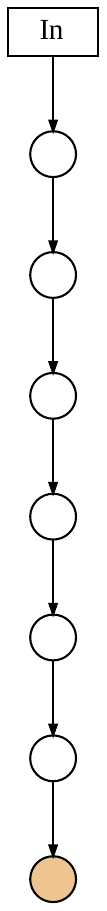}
        \qquad
        $\displaystyle
        \begin{bmatrix}
            1 & 0 & 0 & 0 & 0 & 0 \\
            0 & 1 & 0 & 0 & 0 & 0 \\
            0 & 0 & 1 & 0 & 0 & 0 \\
            0 & 0 & 0 & 1 & 0 & 0 \\
            0 & 0 & 0 & 0 & 1 & 0 \\
            0 & 0 & 0 & 0 & 0 & 1
        \end{bmatrix}
        $
    \end{minipage}%
    \begin{minipage}[t]{0.5\textwidth}
        \centering
        \includegraphics[valign=c,width=0.238\linewidth]{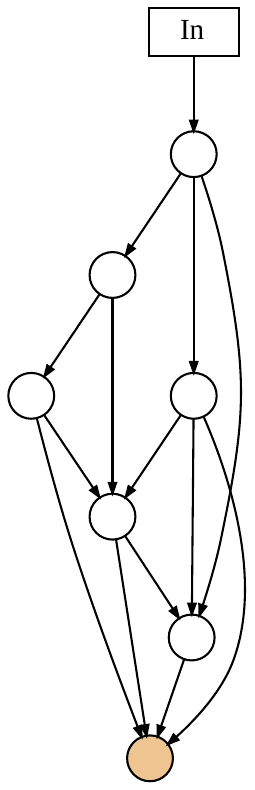}
        \qquad
        $\displaystyle
        \begin{bmatrix}
            1 & 0 & 0 & 0 & 0 & 0 \\
            0 & 1 & 0 & 0 & 0 & 0 \\
            1 & 0 & 0 & 0 & 0 & 0 \\
            0 & 1 & 1 & 1 & 0 & 0 \\
            1 & 0 & 0 & 1 & 1 & 0 \\
            0 & 0 & 1 & 1 & 1 & 1
        \end{bmatrix}
        $
    \end{minipage}
    \caption{Graph representations of the six layer MLP (left) and six layer RandDense network (right) from Figure \ref{fig: dense and rand dense} shown side-by-side with their respective adjacency matrices.}
    \label{fig: dense and rand dense adj mats}
\end{figure*}

\textbf{Network connectivity.} Since the layers within our RandDense networks are randomly wired, the first step in generating such a network is determining which layers are connected to which, or the network connectivity. Following \citet{geiss2022emulating}, given a fixed number of layers $n$, we randomly select the number of neurons per dense layer and generate an adjacency matrix representing the connections between layers in our RandDense network. Since our RandDense networks are still feedforward in nature, several constraints may be placed on the adjacency matrix. If we let each row represent a layer, and column values represent connections (inbound edges) from previous layers, then the adjacency matrix must be lower triangular. Thus, there are $n(n+1)/2$ possible connections for an $n$ layer random network. For each randomly wired model, we generate an adjacency matrix from these possibilities by randomly assigning 1's and 0's to each entry of an $n \times n$ lower triangular matrix. We do not do any further filtering or selection on the connectivity patterns within the randomly wired models. The corresponding adjacency matrices for the networks in Figure \ref{fig: dense and rand dense} are shown in Figure \ref{fig: dense and rand dense adj mats}. Some of these connectivity patterns may have layers without an inbound or outbound tensor. Thus, to ensure the network is valid, we follow \citet{geiss2022emulating} and iterate through each row and column, randomly activating an edge in each row/column if it has no active edges.

\textbf{Node operations.} Nodes in a RandDense network may have multiple inbound or outbound edges, unlike the nodes of an MLP which only have one of each (see Figure \ref{fig: dense and rand dense}). To treat this difference, we must define a new operation that occurs at every node.

The node operation of our RandDense networks begins with an aggregation of the incoming tensors. If a node has more than one inbound edge, we follow \citet{xie2019exploring} and aggregate the incoming tensors via a weighted sum with learnable, positive weights. The weights are kept positive by applying the sigmoid function. Exclusively summing inbound tensors instead of concatenating them \citep{geiss2022emulating} requires a consistent number of neurons per layer throughout the network, but avoids extremely large input tensors that might occur using concatenation. In the style of \citet{xie2019exploring}, our random networks have the ReLU activation function \textit{before} the dense layer so that the outbound tensor can contain both positive and negative values. This avoids extremely large weighted sums when the number of inputs to a given layer is high. Figure \ref{fig: dense and rand dense} illustrates the differences between a node, which contains the dense layer and its activation, in a standard MLP and our random networks. Notice that a node in the graph representation of an MLP in Figure \ref{fig: dense and rand dense} contains a dense layer followed by ReLU activation, while a node of the RandDense network in Figure \ref{fig: dense and rand dense} contains an weighted summation followed by ReLU activation then the dense layer. Lastly, if a node is connected to multiple downstream nodes, it will simply send out identical copies of its output along each outbound edge.

\textbf{Input and output nodes.} In our implementation of RandDense networks, each hidden layer has the same number of neurons. To accommodate smaller input vectors, the node immediately following the input layer will contain a number of neurons equal to the difference between the selected layer size and the input size. For example, if the selected layer size is 100 and the input size is 12, the first randomly wired node following the input layer will contain a dense layer of size 88. Then, this dense layer is concatenated with the input to create a layer of the correct size. This is done so that any nodes downstream of the first node may still have direct access to the input \citep{geiss2022emulating}.

For the output node which may have multiple inbound edges, we simply take the average of all inbound tensors and send this value to a final dense layer of the desired output size. This final output node, which is colored brown in Figure \ref{fig: dense and rand dense}, has linear activation instead of ReLU so that both positive and negative values may be output. Our randomly wired neural networks are implemented using Tensorflow 2 in Python \citep{tensorflow2015-whitepaper}, with code available on Github (see data availability statement).

\section{Data}\label{sec: data}
Standard benchmarks are a valuable tool for the intercomparison of ML methods. The ClimateBench dataset, along with its associated evaluation criteria and metrics seeks to provide a standardized framework for objectively evaluating ML-driven climate model emulators \citep{watson2022climatebench}. We use this ClimateBench dataset to train, validate, and test our models in this work. The dataset is constructed from several simulations performed by Norwegian Earth System Model version 2 (NorESM2) \citep{seland2020overview} as part of the Scenario Model Intercomparison Project (ScenarioMIP) \citep{oneill2016scenario}, Coupled Model Intercomparison Project Phase 6 (CMIP6) \citep{eyring2016overview}, and Detection and Attribution Model Intercomparison Project (DAMIP) \citep{gillett2016detection}. ClimateBench provides four main anthropogenic forcing agents as predictors: carbon dioxide (CO$_2$), sulfur dioxide (SO$_2$), black carbon (BC), and methane (CH$_4$), with the goal of predicting the annual mean surface air temperature (TAS), annual mean diurnal temperature range (max TAS $-$ min TAS) (DTR), annual mean total precipitation (PR), and 90th percentile of daily precipitation (PR90) from 2015-2100. SO$_2$, BC, and the predictands are provided as annual mean spatial distributions across 96 latitude $\times$ 144 longitude global grid points, while the longer lived and well mixed CO$_2$ and CH$_4$ inputs are provided as annual global total concentration and global average emissions, respectively. For each experiment, the ClimateBench dataset also includes the post-processed output of three NorESM2 ensemble members which sample internal variability of the model using different initial model states.

Our models are trained to predict one of TAS, DTR, PR, and PR90 following the ClimateBench benchmarking framework. Specifically, model inputs are global total concentrations of CO\textsubscript{2}, global average CH\textsubscript{4} emissions, as well as spatial distributions ($96 \times 144$) of annual average BC and SO\textsubscript{2} for a range of years from a given set of experiments. The models predict an annual average output variable value for each of the $96 \times 144$ grid points for each of the years 1850-2100, though they are evaluated on a smaller window as specified in the following subsections. 

\subsection{Training and Validation}\label{subsec: train val}
Following the training, validation, and testing strategy in the original ClimateBench manuscript, we select the historical, ssp126, ssp370, and ssp585 experiments from CMIP6 \citep{eyring2016overview} as well as the hist-GHG and hist-aer experiments from DAMIP \citep{gillett2016detection} for training. Each historical experiment spans the years 1850-2014, and each of the ssp experiments contain data for the years 2015-2100 for a total of 753 years. These experiments together cover a wide range of values for each of the input anthropogenic forcers and predicted output variables, making it an ideal suite to train our models on.

In any machine learning application, it is good practice to reserve a portion of the training data for validation so that model performance may be monitored on both sets to prevent overfitting. Since climate data is highly correlated in time, it is recommended to select a continuous portion of the dataset as validation, rather than a random subset. We choose to use the first two years of every decade from the historical, hist-GHG, hist-aer, ssp126, ssp370, and ssp585 training datasets as validation. Here, the validation set is used for early-stopping, a form of regularization where training is halted once performance on the validation set stops improving. 

\subsection{Testing and Evaluation}\label{subsec: test eval}
In addition to the validation set, a third test dataset is used as a final evaluation to ensure that neural network hyperparameter tuning or model selection has not overfit the validation data. As in the original ClimateBench manuscript, we select the ssp245 experiment from ScenarioMIP for the years 2080-2100 as our test dataset. This experiment lies between the extremes of the ssp126 and ssp585 experiments and represents a medium forcing and mitigation scenario.

Following the original work by \citet{watson2022climatebench}, we choose to evaluate our models using ``total root-mean square error'' ($NRMSE_t$), a combination of the normalized, global mean RMSE ($NRMSE_s$) and the NRMSE in the global mean ($NRMSE_g$). These are calculated as
\begin{align*}
    NRMSE_s &= \sqrt{\langle\left(\left|x_{i,j,t}\right|_t-\left|y_{i,j,n,t}\right|_{t,n}\right)^2\rangle} \Big/ \left|\langle y_{i,j}\rangle\right|_{t,n}, \\
    NRMSE_g &= \sqrt{\left|\left(\langle x_{i,j,t}\rangle -\langle\left|y_{i,j,n,t}\right|_n\rangle\right)^2\right|_t} \Big/ \left|\langle y_{i,j}\rangle\right|_{t,n}, \\
    NRMSE_t &= NRMSE_s + \alpha \cdot NRMSE_g,
\end{align*}
where the global mean denoted $\langle x_{i,j} \rangle$ is cosine weighted by latitude to account for the decreasing grid-cell area towards the poles such that $\langle x_{i,j} \rangle = \frac{1}{N_{lat}N_{lon}}\sum_i^{N_{lat}}\sum_j^{N_{lon}}\cos(lat(i))x_{i,j}$, $|x|_t$ is the average of $x$ over all $t$, $n$ is the NorESM2 ensemble members, $t$ is time in years, and $\alpha$ is an empirically chosen factor to give each NRMSE component equal weight \citep{watson2022climatebench}.

\section{Experimental Setup}\label{sec: exp setup}
\begin{figure*}[t]
    \centering
    \includegraphics[width=\linewidth]{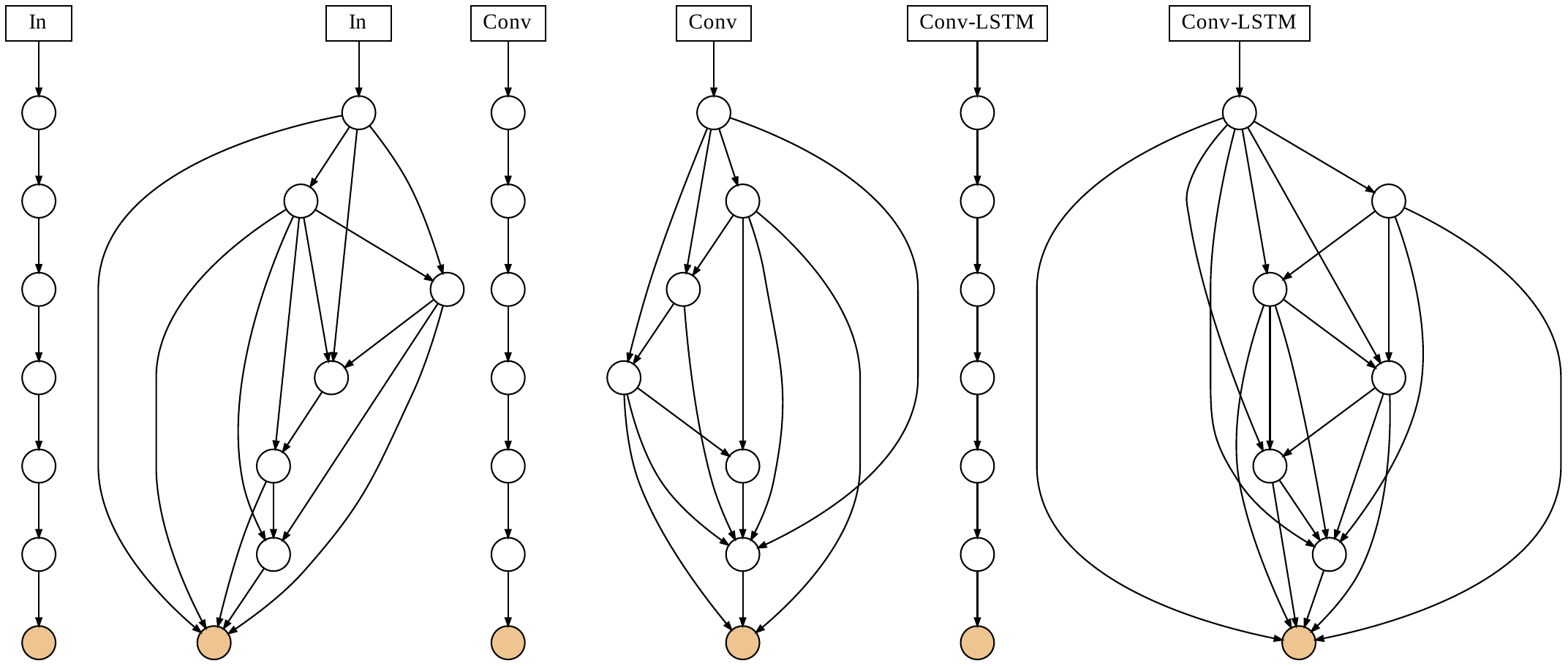}
    \caption{Each of the three baseline models with six MLP dense layers shown next to an example RandDense variation.}
    \label{fig: network examples all}
\end{figure*}

In order to explore the effects of random wiring, we first define three baseline model architectures which contain MLP dense layers. These models are meant to represent realistic deep learning architectures a climate scientist would use for emulation tasks, and include a standard MLP, convolutional neural network, and a convolutional-LSTM network. For each of these model architectures, we explore performance differences across the four output variables (TAS, DTR, PR, and PR90) described in Section \ref{sec: data} at two different parameter counts within the dense layers of each respective network: 1 million and 10 million. With each parameter count, we also explore performance for nine different numbers of hidden dense layers (2-10). Then, the standard MLP dense layers are replaced with randomly wired ones for comparison. We expect that the three baseline model architectures should cover a wide range of performances, illustrating how random wiring impacts performance when placed within both good and bad models. Example networks (with six hidden layers) for each of the three baseline models along with example randomly wired variations, are shown in Figure \ref{fig: network examples all}.

\subsection{MLP Network}\label{subsec: exp setup dense}
A standard MLP model is the most common and basic type of neural network which only contains dense layers. They have been extensively used in climate modelling tasks such as subgrid process representation \citep{rasp2018deep}, rainfall downscaling \citep{tran2019downscaling}, longwave raditaion emulation \citep{krasnopolsky2005new}, pan evaporation prediction \citep{ghorbani2018pan}, and stochastic synthesis in hydrology simulations \citep{rozos2021multilayer}. 

We begin with the first parameter limit of 1 million. We generate MLP models for a given number of hidden layers by selecting a fixed layer size for all of the hidden layers such that the network's parameter count is 1 million $\pm 10\%$. We generate 50 such MLP models with 2 hidden dense layers, 50 MLPs with 3 hidden dense layers, and so on up to 10 hidden dense layers. This gives 450 MLP models, each with different parameter initializations. The process is repeated for a parameter limit of 10 million, yielding a total of 900 MLPs. For the RandDense comparison networks, we repeat a similar process. Using the network generation method described in Section \ref{sec: rand nn}, we generate 50 RandDense networks for 2-10 hidden layers at both the 1 million and 10 million parameter count for a total of 900 RandDense networks. For small numbers of hidden layers, many of these models will have identical layer connectivity patterns since there are very few pattern choices. However, for larger numbers of hidden layers, many of the generated models will have unique wirings.

As discussed in Section \ref{sec: data}, the inputs to the models are global CO$_2$ and CH$_4$, as well as $96 \times 144$ grids of SO$_2$ and BC. However, in MLPs each input is handled by one neuron in the input layer. As such, if we directly fed the anthropogenic forcer inputs to the model, the input layer would be over 26,000 neurons wide. This is a extremely high layer size, and so we follow \citet{watson2022climatebench} and perform dimensionality reduction on the SO$_2$ and BC inputs. Specifically, we use the first five empirical orthogonal functions (EOFs) of each field. Thus, the total input size to the MLP is 12: the first five EOFs for SO$_2$ and BC, plus global CO$_2$ and CH$_4$. The output of the MLP is a flattened map of the predicated variable, which is then reshaped to the correct grid for final predictions.

Both the MLPs and RandDense networks are trained with a batch size of 25 and use the Adam optimizer \citep{kingma2014adam} and mean squared error loss for 100 epochs. We also use early stopping during training with a patience of 10 epochs.

\subsection{Convolutional Network}\label{subsec: exp setup cnn}
Convolutional neural networks (CNNs) have been widely adopted in object tracking and image recognition tasks because they are able to model spacial dependencies \citep{li2021survey,o2015introduction}. CNNs have also been recently adopted for various weather and climate tasks such as precipitation nowcasting \citep{trebing2021smaat}, bow echo detection \citep{mounier2022detection}, climate zone classification \citep{yoo2019comparison} and ocean modelling \citep{nikolaev2020deep}.

Following a similar process as in the previous subsection, we generate 50 MLPs and 50 RandDense networks with 2-10 hidden dense layers for both 1 and 10 million parameter counts. However, rather than being standalone networks, these models are appended to a convolutional block. This block consists of a single convolutional layer with 20 filters, a kernal size of 3, ReLU activation, and $L_2$ regularization \citep{cortes2012l2}. The convolutional layer is followed by global average pooling.

Since CNNs are designed to handle images with multiple channels, we preserve the original dimensionality of the input variables, unlike the transformation we conducted for the standalone MLP and RandDense networks in the previous subsection. Specifically, we are able to feed the $96 \times 144$ maps of SO$_2$ and BC to the convolutional network, as well as global CO$_2$ and CH$_4$. In order to treat the four input variables as four channels of one $96 \times 144$ ``image" of the globe, the CO$_2$ and CH$_4$ data are transformed to $96 \times 144$ grids where each grid cell has the same value.

The CNNs are followed by either conventional MLP or RandDense blocks, which we will simply refer to as ``CNN networks,'' and ``CNN RandDense networks.'' The CNNs use a batch size of 25 and are trained with the Adam optimizer using the mean squared error loss function for 100 epochs. We again use early stopping during training with a patience of 10 epochs.

\subsection{Convolutional-LSTM Network}\label{subsec: exp setup cnn-lstm}
Long short-term memory (LSTM) networks are a type of recurrent neural network (RNN) which model temporal dependencies. This time-aware property makes them useful for a range of forecasting tasks such as precipitation downscaling \citep{tran2019downscaling}, soil moisture estimation \citep{ahmed2021lstm}, and flood forecasting \citep{liu2020applicability}. For the specific task of climate model emulation, a combined CNN-LSTM model has been shown to outperform CNN and LSTM models in isolation \citep{watson2022climatebench}.

As in the previous two subsections, for this architecture we generate 50 MLPs and 50 RandDense networks with 2-10 hidden layers for 1 and 10 million parameter counts. These models are then appended to a CNN-LSTM block. This block first consists of the same convolutional, average pooling, and global average pooling layers as in the previous subsection, but with each of them time distributed (applied in the same way) across every 10 year time window within the training samples. These time distributed layers are followed by an LSTM with 25 units.

In order to enable the CNN-LSTM block to make time-aware predictions, the training data is transformed slightly by slicing it into 10-year moving windows with a 1-year stride. For example, two such windows might be 2015-2024 and 2016-2025. This results in each historical dataset losing 9 data points since the last starting year of the 10-year window is 2005 instead of 2014. Thus, the CNN-LSTM dataset contains $753-3 \cdot 9=726$ data points.

The CNN-LSTMs are followed by either conventional MLP or RandDense blocks, which we will simply refer to as ``CNN-LSTM networks,'' and ``CNN-LSTM RandDense networks.'' We train them with a batch size of 25 using Adam optimizer and mean squared error loss  for 100 epochs. We also use early stopping with a patience of 5 epochs.

\section{Results}\label{sec: results}
\begin{table*}[ht]
\centering
\caption{Best total RMSE performance for each model class and predicted variable across all generated models, along with the original CNN-LSTM and pattern scaling models from \cite{watson2022climatebench}. Lower is better, and the better RMSE between the standard and RandDense models is \textbf{bolded}.}
\begin{tabular}{cccccc}
     & & TAS & DTR & PR & PR90\\
    \toprule
    \multirow{2}{*}{MLP} & Standard & 1.928 & 15.62 & 4.663 & 5.651 \\
    & RandDense & \textbf{1.612} & \textbf{14.67} & \textbf{4.472} & \textbf{5.206} \\
    \midrule
    \multirow{2}{*}{CNN} & Standard & \textbf{3.350} & 23.15 & 9.235 & 10.30 \\
    & RandDense & 3.353 & \textbf{22.92} & \textbf{8.681} & \textbf{9.964} \\
    \midrule
    \multirow{2}{*}{CNN-LSTM} & Standard & \textbf{0.262} & 11.85 & 2.861 & 3.880 \\
    & RandDense & 0.263 & \textbf{11.66} & \textbf{2.775} & \textbf{3.810} \\
    \midrule
    \multirow{2}{*}{ClimateBench} & Neural Network & 0.327 & 16.78 & 3.175 & 4.339 \\
    & Pattern Scaling & 0.320 & 19.72 & 3.662 & 4.461 \\
    \bottomrule
\end{tabular}
\label{tab: total rmse best performance}
\end{table*}
A summary of our experimental results is shown in Table \ref{tab: total rmse best performance}, which shows the best total RMSE ($NRMSE_t$) performance for each model class across all generated models, including all numbers of hidden layers and both parameter counts. For nearly every model architecture and predicted variable, the RandDense variations outperformed their standard counterparts, the two exceptions being the TAS prediction task in both convolutional models. In general, the best performance differences were more stark for the precipitation variables (PR and PR90). Additionally, the margin of difference for instances when the RandDense networks performed better were generally larger than those when the standard networks performed better. A more detailed breakdown is shown in Table \ref{tab: avg total rmse improvement}, which shows the relative total RMSE performance improvement provided by the addition of random wiring. There are a few key trends. First, we see that the performance improvement provided by the addition of random wiring is most stark for MLPs with 1M parameters, but decreases with increasing model architecture complexity with some variety with respect to parameter count. We hypothesize that this is due to the fact that the randomly wired layers in the more complex convolutional architectures are farther removed from the input, and so they do less direct computation on the input as opposed to the CNN or LSTM which directly handle spatial and temporal dependencies in the data directly. Second, there appears to be a more consistent statistically significant benefit provided by random wiring for the precipitation prediction tasks. Lastly, while not every performance change was statistically significant at the $\alpha=0.05$ significance level, there were far more instances of statistically significant performance improvements provided by random wiring than deficits. In fact, there was only one instance (MLPs with 10M parameters predicting DTR) where the addition of random wiring incurred a statistically significant performance loss, compared to the 14 cases where random wiring showed a statistically significant performance improvement.

\begin{table*}[ht]
\centering
\caption{Relative change in best RMSE performance with the addition of random wiring averaged over all generated models for each architecture, parameter count, and predicted variable. Statistically significant changes (p-value < $0.05$) are \textbf{bolded}.}
\begin{tabular}{cccccc}
     & & TAS & DTR & PR & PR90\\
    \toprule
    \multirow{2}{*}{MLP} & 1M & \textbf{{30.4\%}} & \textbf{{12.8\%}} & \textbf{{20.9\%}} & \textbf{{21.8\%}} \\
    & 10M & {5.39\%} & \textbf{{-11.0\%}} & {2.57\%} & {-0.21\%} \\
    \midrule
    \multirow{2}{*}{CNN} & 1M & {-0.10\%} & \textbf{{1.11\%}} & \textbf{{1.06\%}} & {0.41\%} \\
    & 10M & \textbf{{6.36\%}} & \textbf{{2.18\%}} & \textbf{{5.20\%}} & \textbf{{2.94\%}} \\
    \midrule
    \multirow{2}{*}{CNN-LSTM} & 1M & {0.01\%} & {0.76\%} & \textbf{{2.90\%}} & \textbf{{1.54\%}} \\
    & 10M & {-0.89\%} & {1.47\%} & \textbf{{4.47\%}} & \textbf{{3.37\%}} \\
    \bottomrule
\end{tabular}
\label{tab: avg total rmse improvement}
\end{table*}

When compared with the CNN-LSTM model from the original ClimateBench manuscript \citep{watson2022climatebench} (see Table \ref{tab: total rmse best performance}), we find that the CNN-LSTM RandDense models outperform across all prediction tasks, with the greatest improvements for temperature variables. As in \cite{watson2022climatebench}, the CNN-LSTM models also consistently outperform a linear pattern scaling approach which uses independent linear regressions to predict each variable within each grid box. Additional results, including best performances for spatial and global RMSE, can be found in the Appendix. The following subsections are dedicated to discussing the experimental results for each of the baseline model architectures in greater detail.

\subsection{MLP}\label{subsec: results mlp}
\begin{figure*}[ht]
    \centering
    \includegraphics[width=\linewidth]{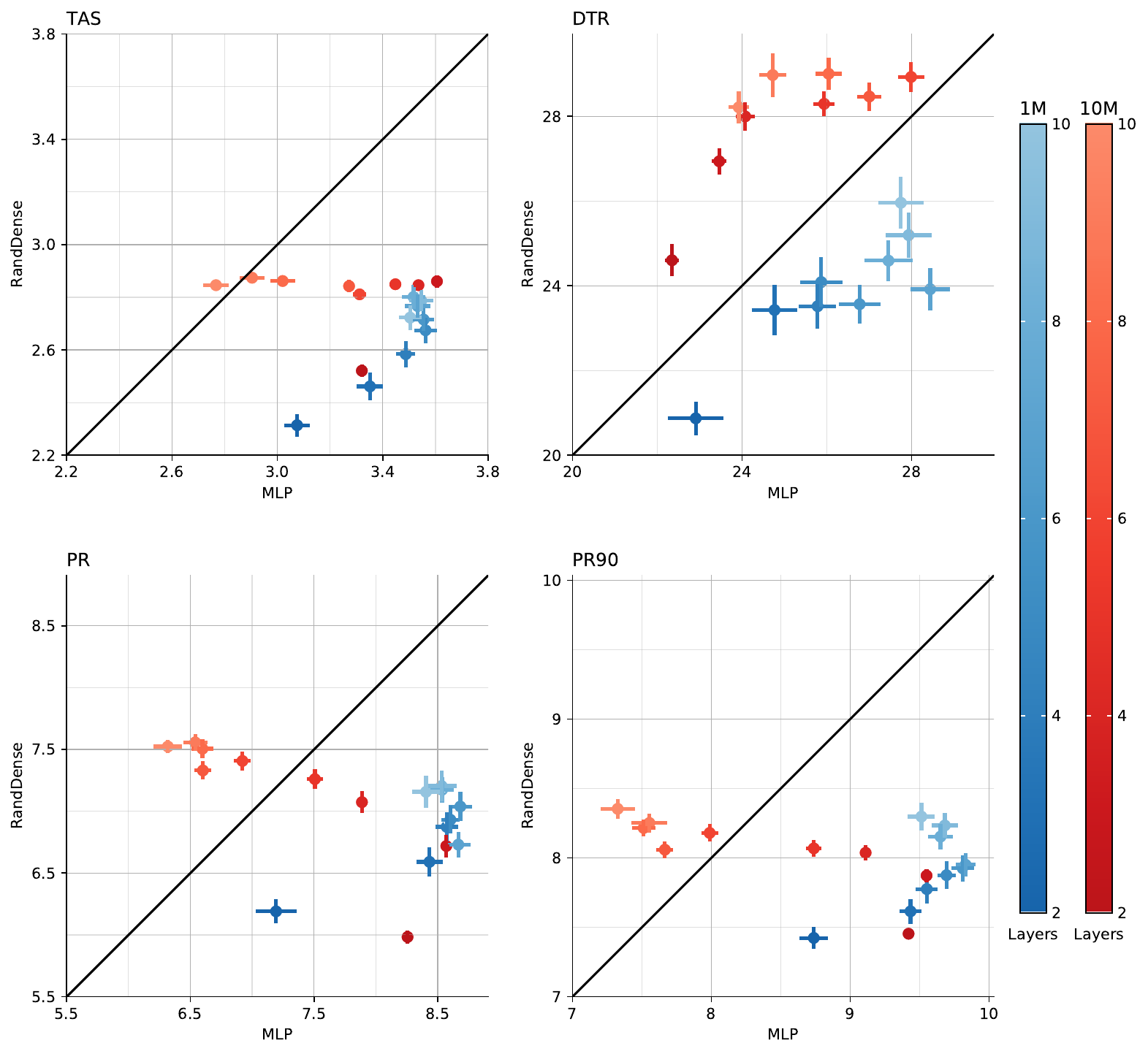}
    \caption{Mean total RMSE of 50 MLP models vs. mean RMSE of 50 RandDense models for both TAS and DTR. The color heatmaps to the right indicate the number of hidden layers. Errorbars show $\pm$ standard error of the mean.}
    \label{fig: mlp total rmse mean}
\end{figure*}
A summary of our MLP and RandDense experiments is shown in Figure \ref{fig: mlp total rmse mean}. Broadly, we find that on average RandDense networks outperform their standard counterparts at 1M parameters with mixed results at 10M, especially for the DTR prediction task. Points below the black $y=x$ line indicate that the RandDense networks outperformed the MLP networks on average, and vice versa for points above. The farther a point is from the $y=x$ line, the more drastic the mean performance difference. For the TAS prediction task, the RandDense networks had better mean RMSE performance across nearly all hidden layer and parameter counts, with the largest mean performance differences being between MLP and RandDense networks with 1M parameters. For the 1M parameter models, both the standard and RandDense variations perform worse with more hidden layers, but at 10M parameters, the trend reverses for standard networks and vanishes for the RandDense networks. For the DTR prediction task, the RandDense models performed better than the MLPs at 1M parameters, and vice versa at 10M parameters. For both parameter counts, mean performance decreases as the number of hidden layers increases. 
Trends for both precipitation variables are quite similar. In general, the RandDense networks outperform their MLP counterparts, with some exceptions for networks with high numbers of hidden layers at 10M parameters. For the MLPs with 1M parameters, performance decreased slightly with higher numbers of hidden layers, but increased slightly with more layers for 10M parameter MLPs. Across both parameter counts, performance for RandDense networks decreased as more layers were added.

\subsection{CNN}\label{subsec: results cnn}
\begin{figure*}[ht]
    \centering
    \includegraphics[width=\linewidth]{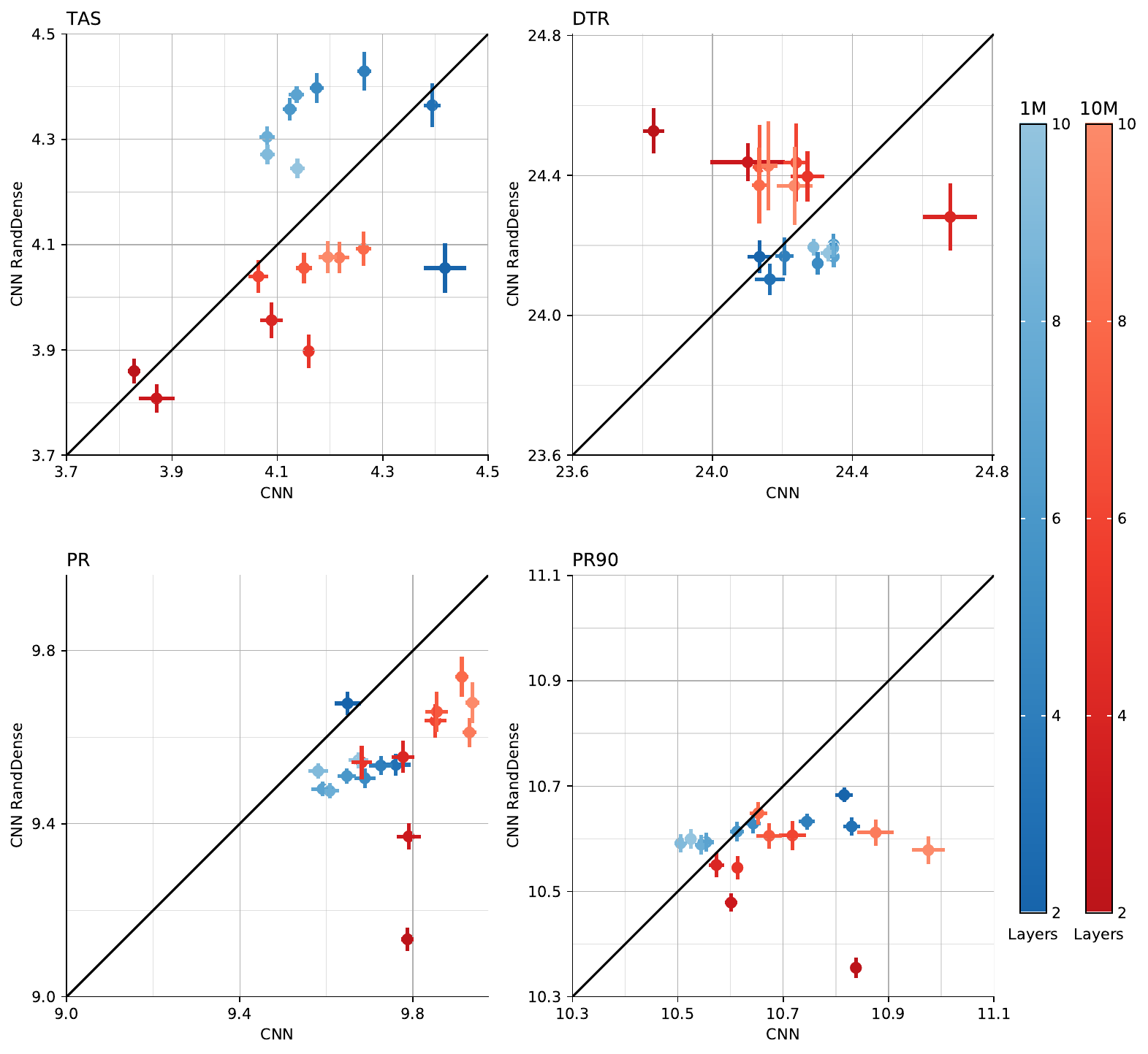}
    \caption{Mean total RMSE of 50 CNN models vs. mean RMSE of 50 CNN RandDense models for both TAS and DTR. The color heatmaps to the right indicate the number of hidden layers. Errorbars show $\pm$ standard error of the mean.}
    \label{fig: cnn total rmse mean}
\end{figure*}
A summary of our CNN and CNN RandDense experiments is shown in Figure \ref{fig: cnn total rmse mean}. In general, the CNN RandDense networks outperform the CNNs for precipitation variables, with mixed results for the temperature variables. For the TAS prediction task, the CNN RandDense networks have overall better mean performance than their standard counterparts at 10M parameters and vice versa at 1M parameters. Interestingly, this pattern is reversed for the DTR prediction task. At 1M parameters, both the CNN and CNN RandDense models had increased TAS mean performance at higher numbers of layers, with the opposite trend at 10M parameters. For DTR, however, neither network architecture showed much change in mean performance with different network depths.

Like the MLP/RandDense networks, the trends in CNN/CNN RandDense mean performance for both precipitation prediction tasks are generally similar. With few exceptions, the CNN RandDense networks outperform their standard counterparts, with the largest mean performance differences at 10M parameters. Furthermore, the mean performance gains for the RandDense models in the PR prediction task are generally greater than in the PR90 prediction task. Mean performance with respect to the number of hidden layers shows no clear trends across prediction variable or number of parameters.

\subsection{CNN-LSTM}\label{subsec: results cnn lstm}
\begin{figure*}[ht]
    \centering
    \includegraphics[width=\linewidth]{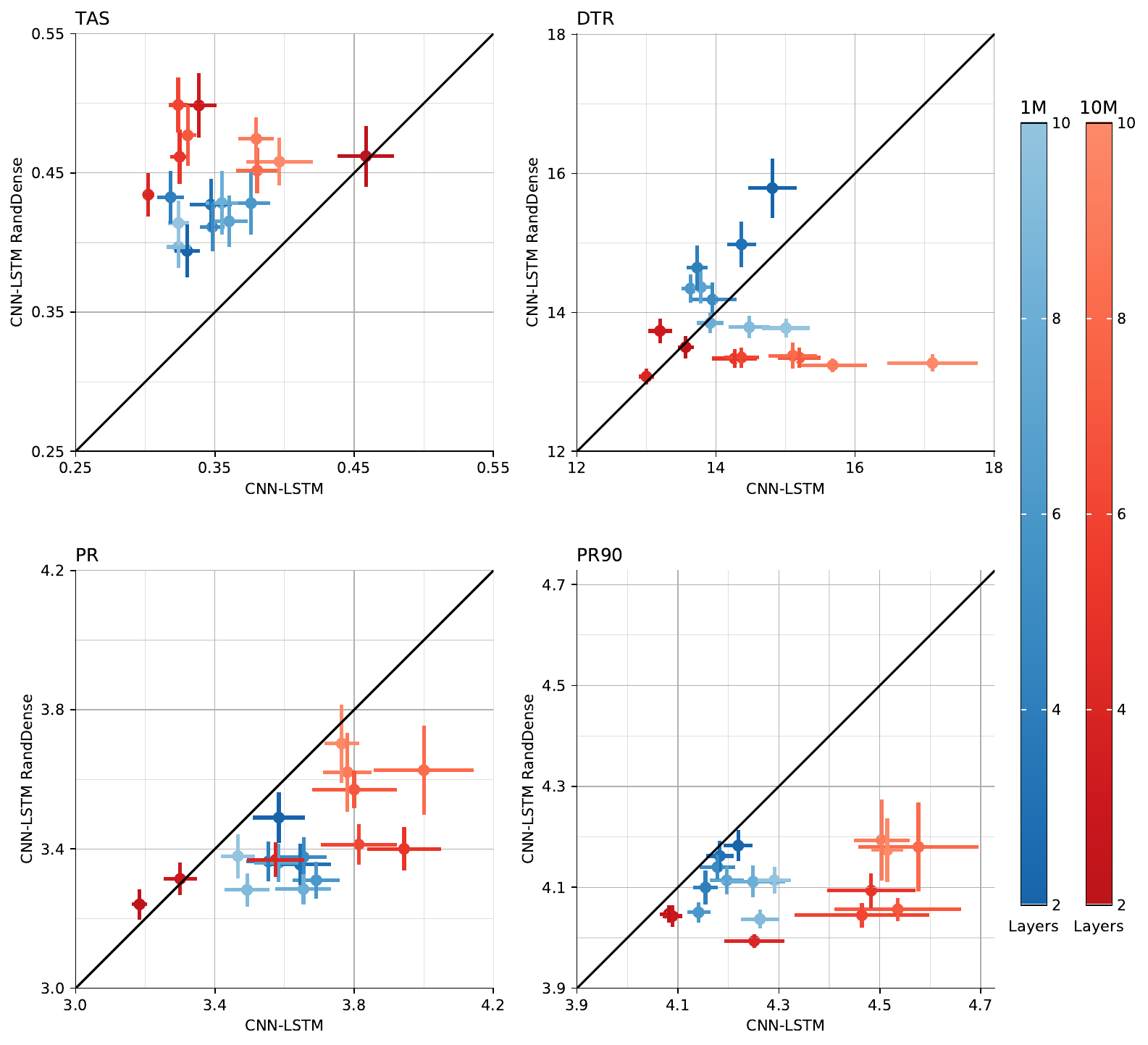}
    \caption{Mean RMSE of 50 CNN-LSTM models vs. mean RMSE of 50 CNN-LSTM RandDense models for both TAS and DTR. The color heatmaps to the right indicate the number of hidden layers. Errorbars show $\pm$ standard error of the mean.}
    \label{fig: cnn lstm total rmse mean}
\end{figure*}
A summary of our CNN-LSTM and CNN-LSTM RandDense experiments is shown in Figure \ref{fig: cnn lstm total rmse mean}. Broadly, the CNN-LSTM RandDense networks outperform, on average, their standard counterparts for precipitation prediction tasks, but show mixed average performance for temperature prediction tasks. Unlike the previously discussed model architectures, the RandDense variations of CNN-LSTM models generally perform worse than their standard counterparts in the TAS prediction task, regardless of network depth or number of parameters. For the DTR prediction task, trends in mean performance differ by parameter count. At 1M parameters, CNN-LSTM models have slightly better mean performance, with the difference becoming less stark as the the number of layers increases. At 10M parameters, the CNN-LSTM RandDense models have generally better mean performance, with larger performance gains at higher numbers of layers.

For both precipitation variables, the CNN-LSTM RandDense networks outperform their standard counterparts with few exceptions across all network depths and parameter counts. Mean performance differences are most stark for the PR90 prediction task at high numbers of layers and 10M parameters. For both PR and PR90, the CNN-LSTM RandDense networks have slight performance gains for deeper networks at 1M parameters, and the slight performance losses for deeper networks at 10M parameters.

\begin{figure*}[ht]
    \centering
    \includegraphics[width=\textwidth]{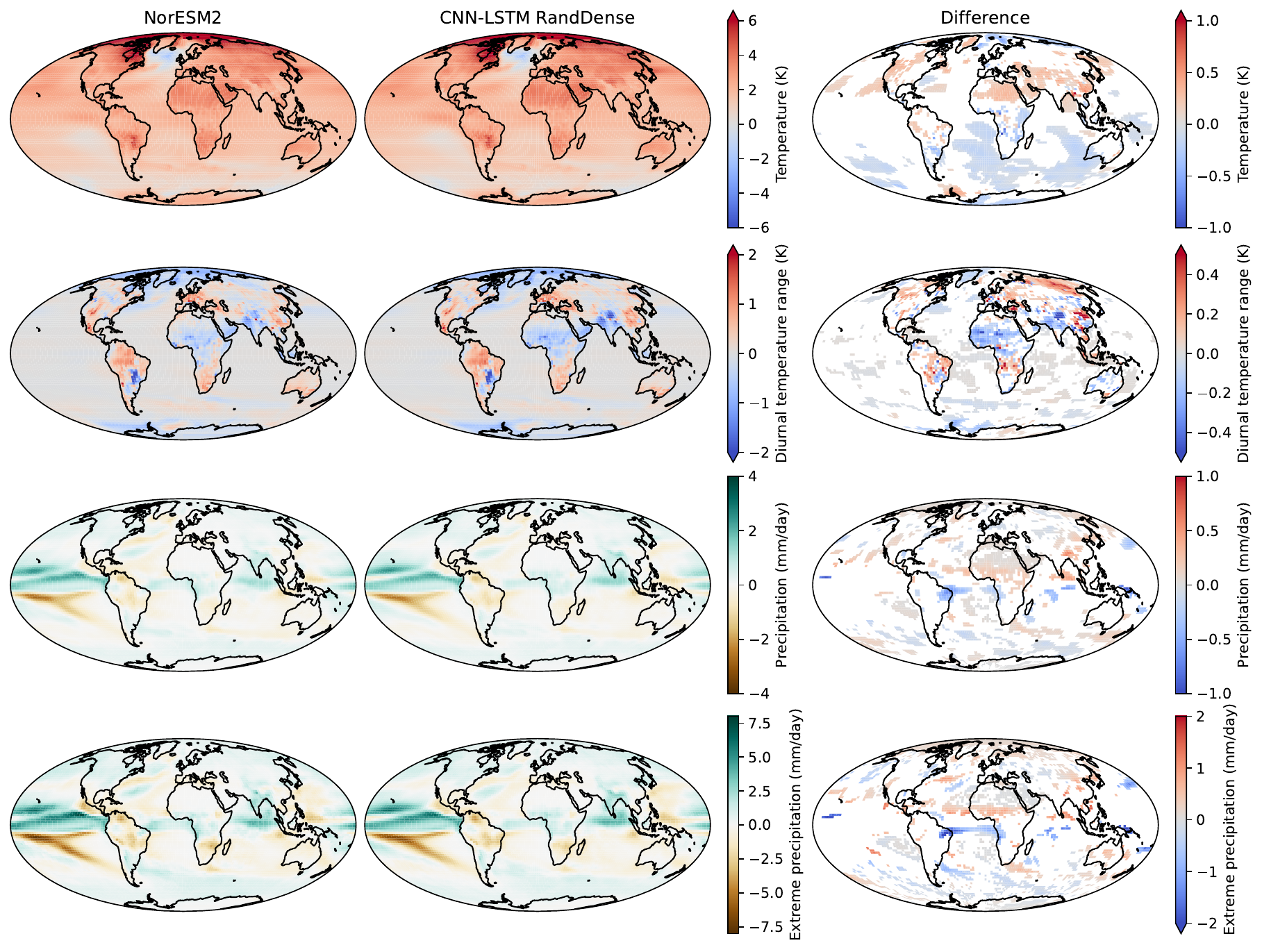}
    \caption{NorESM2 ground truth predictions for each variable averaged from 2080-2100, alongside the predictions averaged over the same time period of the best performing CNN-LSTM RandDense models. Also shown is the mean difference between the NorESM2 and CNN-LSTM RandDense predictions, once again averaged over the 2080-2100 testing period. Statistically insignificant differences $(p>0.05)$ are masked.}
    \label{fig: cnn-lstm rand dense noresm2 comparison}
\end{figure*}

Since the CNN-LSTM RandDense model had the best performance overall (see Table \ref{tab: total rmse best performance}), we further analyze its difference from the NorESM2 ground truth predictions. Figure \ref{fig: cnn-lstm rand dense noresm2 comparison} shows the NorESM2 predictions alongside the best CNN-LSTM RandDense predictions, as well as the mean difference between the two. Across prediction tasks, the mean difference in predictions during the testing period (2080-2100) was statistically insignificant in most locations. Furthermore, significant differences, when present, were relativey small. For example, about 54\% of the significant differences in TAS prediction are within 0.1K ($\sim 4.5\%$ error) and 90\% of significant differences are within 0.22 K ($\sim 11.8\%$ error).

In general, the CNN-LSTM RandDense model is more performant over the ocean than land for temperature variables. This is especially true for DTR. Among statistically significant errors (see Figure \ref{fig: cnn-lstm rand dense noresm2 comparison}), the CNN-LSTM RandDense model had an average DTR error of 0.05K over sea compared to a 0.17K over land. Additionally, it tends to overestimate warming in the Northern hemisphere but underestimates in the Southern hemisphere. For the precipitation variables, the model generally makes better predictions over land. The area which the models makes the worst predictions in is the intertropical convergence zone (ITCZ) between South America and Africa. Specifically, the model underestimates both average and extreme precipitation over the ITCZ's winter location \citep{schneider2014migrations}. We hypothesize that it struggles in this region due to variability in climate change-induced ITCZ shift across the simulations the model was trained on.

\section{Discussion}\label{sec: discussion}
Each model architecture experiment set shares a common theme across the experimental results presented here: there is at least one performance metric and prediction task where randomization provides clear benefits. For example, even though randomization did not provide clear mean DTR performance benefits, Table \ref{tab: total rmse best performance} shows that the best performance metric favored the RandDense models across all tested architectures. This suggests that for the task of climate model emulation, scientists and practitioners are likely to achieve some performance benefit from generating a limited number of randomly wired variations of their existing models with a similar number of parameters. In scenarios where finely-tuned and optimized model predictive skill is of critical importance, investigating network randomization is an important design choice to consider. With this in mind, the remainder of this section is dedicated to further exploring random networks and why they might achieve the results they do.

\subsection{Prediction Speed}\label{subsec: prediction speed}
The speed with which neural networks make their predictions is important for both standalone models and models within a larger system. Researchers seeking to quickly construct, validate, and test standalone models for multiple applications would clearly benefit from faster models. Similarly, researchers seeking to use a neural network to model a subprocess in a large climate model \citep{irrgang2021towards, holder2022using, silva2021physically, geiss2022emulating} would also benefit from quicker predictions, since such submodels are often called many thousands of times per timestep in a  full Earth system model. As such, we investigate the prediction speed of randomly wired neural networks and compare them with their feedforward counterparts.

For fair comparison, we took the a best performing standard model, say a 6 layer MLP, and the best performing RandDense equivalent and used them to make predictions on the test data 10 times. This is done because the time required for just one prediction is quite small. We repeated nine more times for a total of ten trials. Lastly, we conducted an independent sample t-test on the two samples of 10 run times to investigate whether the mean run times of the two models were different in a statistically significant way.

\begin{figure*}[ht]
    \centering
    \includegraphics[width=\textwidth]{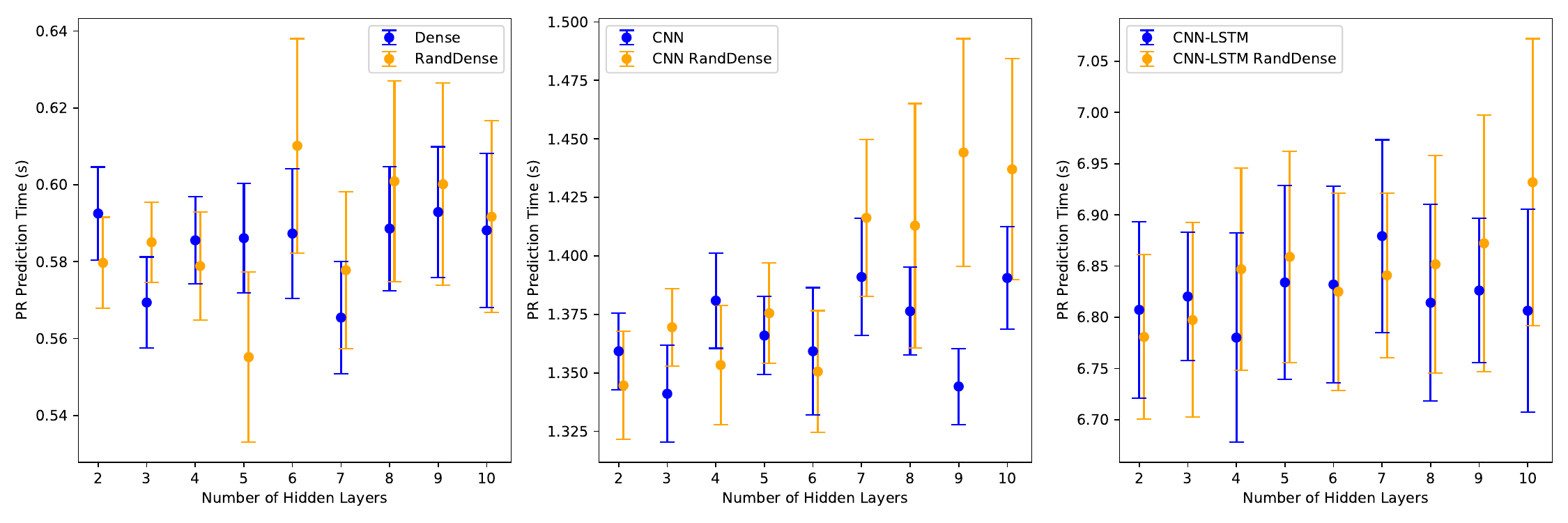}
    \caption{The average PR prediction times for the best performing models within each model class along with 95\% confidence intervals.}
    \label{fig: pr prediction speeds}
\end{figure*}

Across the several different model architectures, layer counts, and prediction tasks we find no consistent difference between the run times of the standard models and their randomly wired variations. In a large majority of comparisons, the difference was statistically insignificant. This is partially illustrated in Figure \ref{fig: pr prediction speeds}, which shows 95\% confidence intervals for PR prediction speed across model architectures. Notice that in nearly every case, the confidence intervals for the standard and RandDense models overlap. Furthermore, neither the standard or RandDense networks consistently had faster average runtimes. We conclude that random wiring neither benefits nor detracts from prediction speed, making RandDense networks suitable replacements for standard MLPs either in standalone or sub-model applications.

\subsection{Sensitivity to the Node Operation}\label{subsec: node op}
The node operation of the random networks discussed in Section \ref{sec: rand nn} and illustrated in Figure \ref{fig: dense and rand dense} was a specific design choice by the authors, inspired by the choices of \citet{xie2019exploring} and \cite{geiss2022emulating}. There are many valid alternative designs, however. Here, we investigate two modifications to our node operation: 
\begin{enumerate}
    \item Placing the ReLU activation after the Dense layer instead of before.
    \item Replacing the weighted sum with a simple unweighted sum.
\end{enumerate}

With the ReLU activation after the dense layer, we found different results across the three model architectures. 1M parameter RandDense network performance decreased for the all prediction tasks by about $20\%$, but in many cases the RandDense networks still outperformed their standard counterparts. The same was observed for 10M parameter networks, except when predicting DTR, where RandDense performance improved by $20\%$. For the CNN RandDense models, there were mixed mean performance changes for the temperature variables, just a few percentage points increase at 1M parameters and the opposite at 10M parameters. There was little change in performance ($<1\%$) predicting precipitation. Lastly, for the CNN-LSTM RandDense models, when the ReLU activation was placed after the dense layer there was a $6\%$ improvement TAS prediction performance, $8\%$ decrease in DTR prediction performance, and limited change in performance ($<1\%$) for both precipitation variables.

Removing the weighted sum and replacing it with a simple unweighted sum made no significant performance difference across all model types, predicted variables, layer counts, or parameter counts. These performance changes, or lack thereof, as a result of node operation modifications highlight the importance of specific components of the operation. 

\subsection{Further Investigating the Effects of Random Wiring}\label{subsec: analysis experiments}
We hypothesize that random wiring may affect a neural network's performance by either
\begin{enumerate}
    \item Benefiting/hindering the training of the full model, or
    \item Processing the output of the previous layers better/worse.
\end{enumerate}
To explore these hypotheses, we attempted to isolate the source of performance benefits or deficits for randomly wired neural networks. Specifically, for the CNN and CNN-LSTM architectures we trained the best performing conventional models from scratch. We then isolated the trained CNN or CNN-LSTM block and froze its weights. Lastly, we attached to this block the best performing randomly wired architecture with randomly initialized weights and retrained the model. This is a similar idea to that of previous work which configured artificially evolved nanoparticle systems into Boolean logic gates \citep{bose2015evolution}. Since the CNN or CNN-LSTM block has been frozen, only the weights in the randomly wired layers are trained in this iteration. If the performance of this new model was different from the original random model which had both its convolutional block \textit{and} random layers trained from scratch, then it would indicate that random wiring was beneficial or detrimental to the training process. On the other hand, if the frozen model and the original random model performed the same, it would indicate that the randomly wired layers are helping or hurting the model in the way they process the output of the CNN/CNN-LSTM block.

\begin{figure*}[ht]
    \centering
    \includegraphics[width=0.9\textwidth]{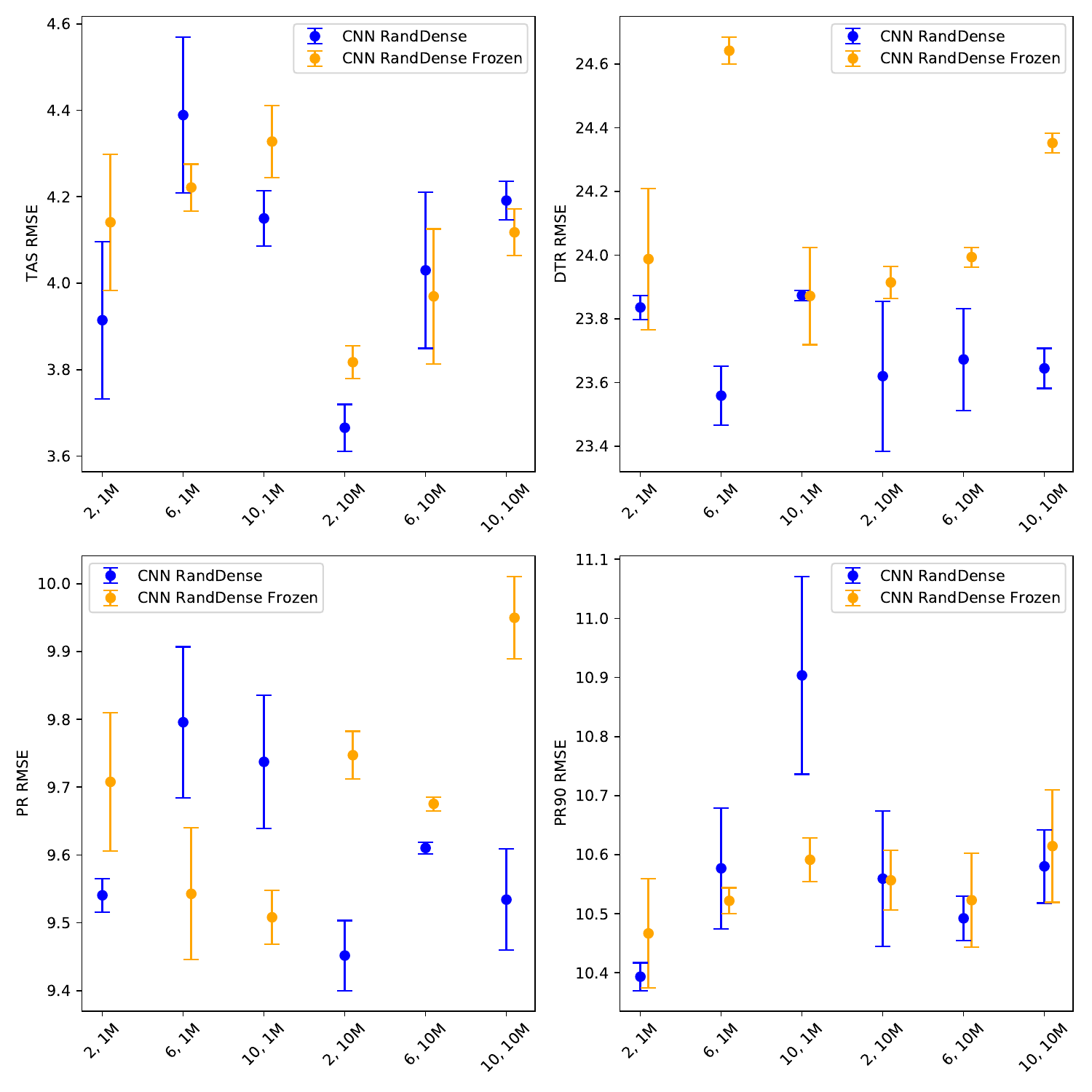}
    \caption{Mean RMSE performance of 10 CNN RandDense networks vs. mean RMSE performance of 10 CNN RandDense networks with frozen weights from the best performing standard CNN network. The x-axis indicates the number of hidden layers and parameters. For example, ``6, 1M" means 6 hidden layers and 1M parameters}
    \label{fig: cnn weight freezing}
\end{figure*}

We repeated this experimental procedure 10 times for both the CNN and CNN-LSTM, analyzing both 1 million and 10 million parameter counts at 2, 6, and 10 layers. All four prediction tasks were analyzed for this experiment. Figure \ref{fig: cnn weight freezing} shows a summary of the results for the CNN RandDense networks. The figure illustrating the CNN-LSTM RandDense results may be found in the Appendix. Overall, we find no consistent trend in performance differences between the RandDense models trained from scratch and those with frozen CNN or CNN-LSTM blocks with only one exception: the DTR prediction task for CNN networks. While the confidence intervals for RandDense networks and their frozen counterparts may not always overlap, neither performs better in general. This provides evidence that random wiring impacts performance due to the way it handles the output of the convolutional block, and not by it affecting the CNN's training process.

\section{Conclusion}\label{sec: conclusion}
Motivated by the promise of randomly wired neural networks in related applications, we explore random wirings between dense layers of neural networks for the task of climate model emulation. We replaced the traditional feedforward dense layers in several model architectures with our randomly wired networks, coined ``RandDense'' networks, and conduct performance experiments using the ClimateBench dataset. Across several architectures, model complexities, and predicted variables, we find performance benefits for models containing randomly wired layers, indicating that in many cases, standard feedforward networks may be effectively replaced with RandDense networks which achieve better performance at no additional computational cost.

Additionally, we explore various node operations for RandDense networks and conduct preliminary experiments to understand their performance (Section \ref{sec: discussion}). However, this leaves the door open for future work. Many more node operation modifications, such as the inclusion of batch normalization and various activation functions, remain to be evaluated. Furthermore, our method of generating random graph structures described in Section \ref{sec: rand nn} is only one way of accomplishing the task. Special defined types of random graphs have been explored in other applications \citep{xie2019exploring}, but assessing their efficacy for climate model emulation remains a topic for future work. Lastly, while the  experiments in Section \ref{sec: discussion} provide a brief exploration into why RandDense networks perform better or worse in certain scenarios, there may be other reasons why RandDense networks perform differently than their standard counterparts such as alleviating the vanishing gradient problem \citep{orhan2017skip}. Further analysis into RandDense networks may shed light on additional reasons behind their performance discrepancies.

%

%

\clearpage
\acknowledgments
The authors would like to thank Joseph Hardin for helpful discussions and feedback. DWP acknowledges funding from the European Union's Horizon 2020 research and innovation programme iMIRACLI under Marie Skłodowska-Curie grant agreement No 860100.

%
%
\datastatement
The  ClimateBench data is available here: \url{https://doi.org/10.5281/zenodo.5196512}. The code used to generate, train, and evaluate our models can be found here: \url{https://github.com/yikwill/randomly-wired-nn}.

%

\appendix\label{appendix}



\begin{figure*}[ht]
    \centering
    \includegraphics[width=0.9\textwidth]{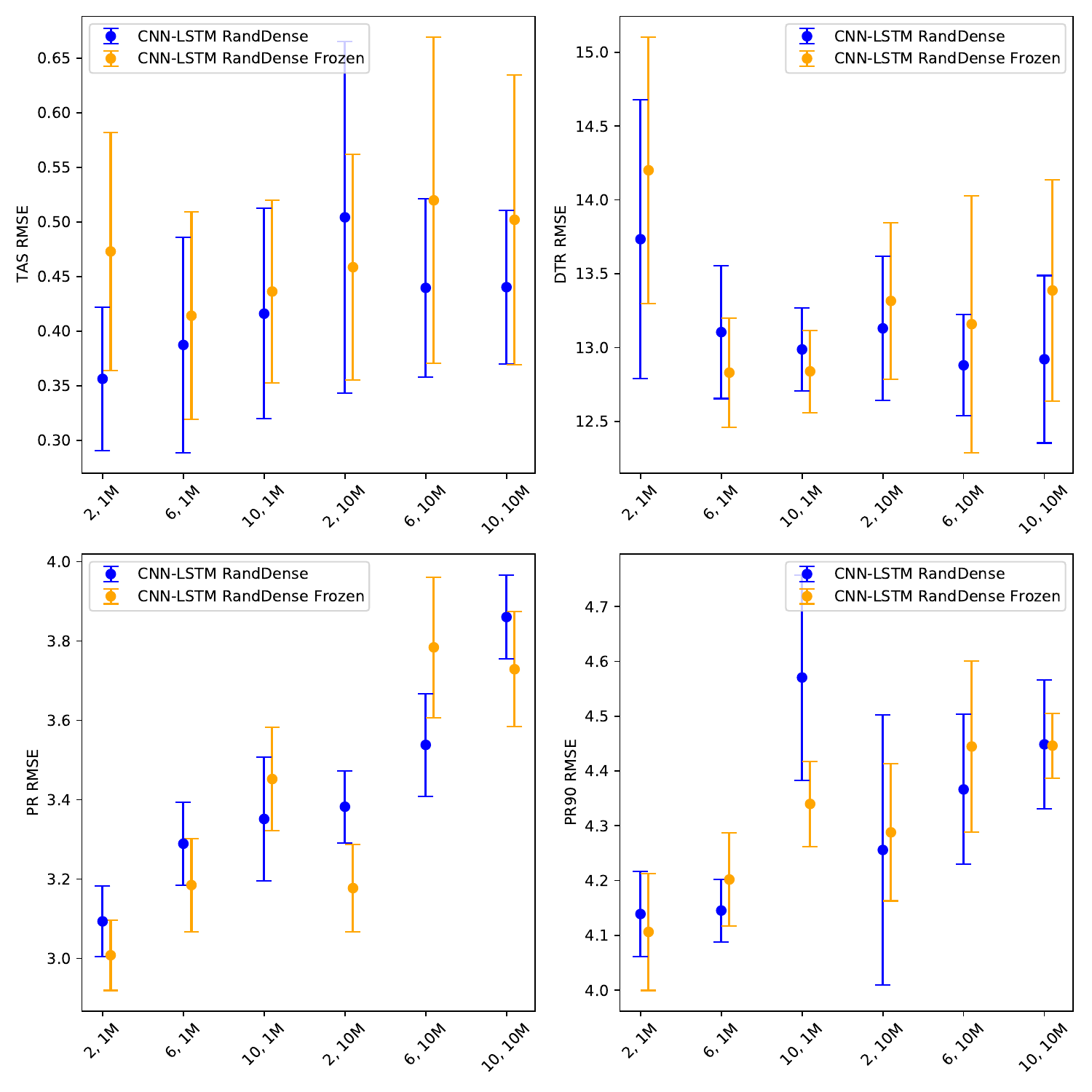}
    \caption{Mean RMSE performance of 10 CNN-LSTM RandDense networks vs. mean RMSE performance of 10 CNN-LSTM RandDense networks with frozen weights from the best performing standard CNN-LSTM network. The x-axis indicates the number of hidden layers and parameters. For example, ``6, 1M" means 6 hidden layers and 1M parameters}
    \label{fig: cnn lstm weight freezing}
\end{figure*}

\appendixtitle{Additional Results}
In this section, we present additional results from our RMSE and analytical experiments. The Tables \ref{tab: all rmse best performance 1M} and \ref{tab: all rmse best performance 10M} show the best spatial, global, and total RMSE performance of each model at each parameter count tested. Figure \ref{fig: cnn lstm weight freezing} shows the results of the weigh freezing analysis experiment for the CNN-LSTM RandDense architecture.

\begin{table*}[ht]
\centering
\caption{Best spatial, global, and total RMSE performance for each model class and predicted variable across all network depths at 1M parameters, along with the original CNN-LSTM model from \cite{watson2022climatebench}. Lower is better, and the better RMSE between the standard and RandDense models is bolded.}
{\def\arraystretch{1.5}
\resizebox{\textwidth}{!}{\begin{tabular}{cccccccccccccc}
    & & \multicolumn{3}{c}{TAS} & \multicolumn{3}{c}{DTR} & \multicolumn{3}{c}{PR} & \multicolumn{3}{c}{PR90} \\
    \cmidrule(lr){3-5} \cmidrule(lr){6-8} \cmidrule(lr){9-11} \cmidrule(lr){12-14}
    & & \rotatebox{90}{Spatial} & \rotatebox{90}{Global} & \rotatebox{90}{Total} & \rotatebox{90}{Spatial} & \rotatebox{90}{Global} & \rotatebox{90}{Total} & \rotatebox{90}{Spatial} & \rotatebox{90}{Global} & \rotatebox{90}{Total} & \rotatebox{90}{Spatial} & \rotatebox{90}{Global} & \rotatebox{90}{Total} \\
    \toprule
    \multirow{2}{*} {MLP} & Standard & 0.433 & 0.380 & 2.331 & 10.76 & \textbf{0.782} & 15.62 & \textbf{2.655} & \textbf{0.188} & \textbf{4.663} & 4.137 & 0.456 & 6.741 \\
    & RandDense & \textbf{0.301} & \textbf{0.262} & \textbf{1.612} & \textbf{9.918} & 0.795 & \textbf{14.67} & 2.729 & 0.266 & 4.472 & \textbf{3.249} & \textbf{0.411} & \textbf{5.306} \\
    \midrule
    \multirow{2}{*} {CNN} & Standard & 0.698 & 0.610 & 3.747 & \textbf{16.44} & 1.307 & 23.61 & \textbf{5.078} & 0.811 & 9.235 & \textbf{5.957} & 0.867 & \textbf{10.30} \\
    & RandDense & \textbf{0.697} & \textbf{0.608} & \textbf{3.737} & 16.56 & \textbf{1.273} & \textbf{23.45} & 5.081 & \textbf{0.800} & \textbf{9.117} & 5.999 & \textbf{0.852} & 10.31 \\
    \midrule
    \multirow{2}{*} {CNN-LSTM} & Standard & \textbf{0.055} & \textbf{0.041} & \textbf{0.262} & \textbf{7.376} & \textbf{0.802} & 12.18 & 1.948 & \textbf{0.169} & 2.861 & 2.346 & \textbf{0.302} & 3.880 \\
    & RandDense & 0.058 & \textbf{0.041} & 0.263 & 7.572 & 0.812 & \textbf{12.13} & \textbf{1.875} & \textbf{0.169} & \textbf{2.775} & \textbf{2.263} & 0.303 & \textbf{3.810} \\
    \midrule
    {ClimateBench} & & 0.107 & 0.044 & 0.327 & 9.917 & 1.372 & 16.78 & 2.128 & 0.209 & 3.175 & 2.610 & 0.346 & 4.339 \\
    \bottomrule
\end{tabular}}}
\label{tab: all rmse best performance 1M}
\end{table*}

\begin{table*}[ht]
\centering
\caption{Best spatial, global, and total RMSE performance for each model class and predicted variable across all network depths at 10M parameters, along with the original CNN-LSTM model from \cite{watson2022climatebench}. Lower is better, and the better RMSE between the standard and RandDense models is bolded.}
{\def\arraystretch{1.5}
\resizebox{\textwidth}{!}{\begin{tabular}{cccccccccccccc}
    & & \multicolumn{3}{c}{TAS} & \multicolumn{3}{c}{DTR} & \multicolumn{3}{c}{PR} & \multicolumn{3}{c}{PR90} \\
    \cmidrule(lr){3-5} \cmidrule(lr){6-8} \cmidrule(lr){9-11} \cmidrule(lr){12-14}
    & & \rotatebox{90}{Spatial} & \rotatebox{90}{Global} & \rotatebox{90}{Total} & \rotatebox{90}{Spatial} & \rotatebox{90}{Global} & \rotatebox{90}{Total} & \rotatebox{90}{Spatial} & \rotatebox{90}{Global} & \rotatebox{90}{Total} & \rotatebox{90}{Spatial} & \rotatebox{90}{Global} & \rotatebox{90}{Total} \\
    \toprule
    \multirow{2}{*} {MLP} & Standard & \textbf{0.354} & \textbf{0.315} & \textbf{1.928} & \textbf{10.30} & \textbf{1.083} & \textbf{18.41} & \textbf{2.484} & 0.453 & 5.146 & \textbf{3.017} & \textbf{0.496} & \textbf{5.651} \\
    & RandDense & 0.410 & 0.338 & 2.100 & 11.05 & 1.369 & 18.58 & 2.880 & \textbf{0.426} & \textbf{5.094} & 3.472 & 0.511 & 6.704 \\
    \midrule
    \multirow{2}{*} {CNN} & Standard & \textbf{0.641} & \textbf{0.542} & \textbf{3.350} & \textbf{15.94} & 1.211 & 23.15 & 5.117 & 0.811 & 9.481 & 5.936 & 0.884 & 10.398 \\
    & RandDense & 0.643 & \textbf{0.542} & 3.353 & 16.56 & \textbf{1.176} & \textbf{22.92} & \textbf{5.114} & \textbf{0.711} & \textbf{8.681} & \textbf{5.905} & \textbf{0.785} & \textbf{9.964} \\
    \midrule
    \multirow{2}{*} {CNN-LSTM} & Standard & \textbf{0.061} & \textbf{0.041} & \textbf{0.264} & \textbf{7.273} & 0.815 & 11.85 & 2.003 & \textbf{0.169} & 2.911 & 2.342 & \textbf{0.301} & 3.906 \\
    & RandDense & 0.063 & \textbf{0.041} & 0.268 & 7.525 & \textbf{0.795} & \textbf{11.66} & \textbf{1.926} & \textbf{0.169} & \textbf{2.816} & \textbf{2.295} & 0.304 & \textbf{3.821} \\
    \midrule
    {ClimateBench} & & 0.107 & 0.044 & 0.327 & 9.917 & 1.372 & 16.78 & 2.128 & 0.209 & 3.175 & 2.610 & 0.346 & 4.339 \\
    \bottomrule
\end{tabular}}}
\label{tab: all rmse best performance 10M}
\end{table*}

\clearpage

%



\bibliographystyle{ametsocV6}
\bibliography{main}

\end{document}